\documentclass[12pt]{article}
\usepackage{graphicx}
\newcommand{\vc}{\vskip0.5cm\noindent}
\newcommand{\vu}{\vskip1.cm\noindent}
\newcommand{\be}{\begin{equation}}
\newcommand{\ee}{\end{equation}}
\newcommand{\ba}{\begin{eqnarray}}
\newcommand{\ea}{\end{eqnarray}}
\begin{document}
\noindent
{\Large \bf Correlating overrepresented upstream motifs to gene
expression: a computational approach to regulatory element discovery
in eukaryotes.}
\vskip1.cm
\noindent
M. Caselle$^{1}$, 
F. Di Cunto$^2$ and P. Provero$^{3,1,*}$
\vskip1.cm\noindent
$^1$ Dipartimento di Fisica Teorica, Universit\`a di Torino, and INFN, 
sezione di Torino, Torino, Italy.
\vc
$^2$ Dipartimento di Genetica, Biologia e Biochimica, 
 Universit\`a di Torino, Torino, Italy.
\vc 
$^3$ Dipartimento di Scienze e Tecnologie Avanzate, Universit\`a del
Piemonte Orientale, Alessandria, Italy.
\vu
e-mail addresses:\\*
 caselle@to.infn.it, ferdinando.dicunto@unito.it,
provero@to.infn.it 
\vu
$^*$ corresponding author
\newpage\noindent
{\bf \large Abstract}
\vc
{\bf Background}
\vc
Gene regulation in eukaryotes is mainly effected through 
transcription factors
binding to rather short recognition motifs generally located upstream 
of the coding region. We present a novel computational method to identify
regulatory elements in the upstream region of eukaryotic genes. 
The genes  are grouped in sets sharing an 
overrepresented short motif in their upstream sequence. 
For each set,
the average expression level from a microarray experiment  is
determined: If this level is significantly higher or  
lower than the average taken over the whole genome, then the
overerpresented motif shared by the genes in the 
set is likely to play a role in their regulation. 
\vc 
{\bf Results}
\vc
The method was tested by applying it 
to the genome of {\em Saccharomyces cerevisiae}, using the
publicly available results of a DNA microarray experiment, 
in which expression levels for virtually all the genes were measured
during the diauxic shift from fermentation to respiration.
Several known motifs were correctly identified, and a new candidate
regulatory sequence was determined. 
\vc
{\bf Conclusions}
\vc 
We have described and successfully tested 
a simple computational method to identify upstream
motifs relevant to gene regulation in eukaryotes by studying the
statistical correlation between overepresented upstream motifs and
gene expression levels.
\newpage
\noindent
{\bf \large Introduction}
\nopagebreak
\vskip0.5cm\noindent
One of the biggest challenges of modern genetics is to extract
biologically meaningful information from the huge mass of raw data
that is becoming available. In particular, the availability of
complete genome sequences on one hand, and of genome-wide microarray
data on the other, provide invaluable tools to elucidate the
mechanisms underlying transcriptional regulation. The sheer amount of
available data and the complexity of the mechanisms at work
require the development of specific data analysis techniques to identify 
statistical patterns and regularities, that can then be the subject of 
experimental investigation.
\vc
The regulation of gene expression in eukaryotes is known to be mainly
effected through transcription factors
binding to rather short recognition motifs generally located upstream 
of the coding region.
One of the main problems in
studying regulation of gene expression is to identify the motifs that
have transcriptional meaning, and the genes each motif regulates.
\vc
The usual approach to this kind of analysis begins by identifying
groups of co-regulated genes, for example  by applying clustering techniques
to the expression profiles obtained from microarray experiments. One
then studies the upstream sequences of a set of coregulated genes
looking for shared motifs. 
Examples of this approach as applied to {\it S. cerevisiae} are 
Refs. \cite{DeRisi:1997, vanHelden:1998,Tavazoie:1999}.
\vc
In this paper we suggest an alternative
method which somehow follows the inverse route: genes are grouped into 
(non-disjoint) sets, each set being characterized by a short motif which is
overrepresented in the upstream sequence. For each set, the average
expression is computed for a certain microarray experiment, and
compared to the genome-wide average expression from the same
experiment. If a statistically significant difference is found, then
the motif that defines the set of genes is a candidate regulatory
sequence. The rationale for looking for overrepresented motifs is that, in 
many instances, regulatory motifs are known to appear repeated many
times within a relatively short upstream sequence
\cite{vanHelden:1998,Wagner:1997}, so that the number of repetitions
turns out to be much bigger  than what would
be expected from chance alone.
\vc
A somehow related approach, which does not require any previous grouping of
genes based on 
their expression  profiles, was presented in
Ref. \cite{Bussemaker:2001}, 
where the effect of upstream motifs on gene expression levels is
modeled by a sum of activating and inhibitory terms. Experimental
expression levels are then fitted to the model, and statistically
significant motifs are identified. Our approach differs in the
importance given to overrepresented motifs, thus considering activation 
and inhibition as an effect that depends on a threshold number of
repetitions of a motif rather than on additive contributions from all
motifs. Clearly the two mechanisms are far from being mutually
exclusive, therefore we expect the candidate regulatory sites found
with the two methods to significantly overlap. 
\vc
However it is important
to notice that the kind of statistical correlation between upstream
motifs and expression that our algorithm identifies does not depend on 
any special assumption on the functional dependence of expression
levels on the number of motif repetitions, as long as this dependence
is strong enough to provide a significant deviation from the average
expression when enough copies of the motif are present.  
A comparison of our results with those obtained in
Ref. \cite{Bussemaker:2001} is provided in the ``Results and
discussion'' section.
\vu
{\bf \large The method}
\nopagebreak
\vc
In general the motifs with known regulatory function are not
identified with a fixed nucleotide sequence, but rather with sequences where 
substitutions are allowed, or spaced dyads of fixed sequences, etc.
However in this study, in order to test the method while keeping the
technical complications to a minimum, we will limit ourselves to fixed 
short nucleotide sequences, that we call {\it words}. While previous
studies (see {\it e.g} \cite{vanHelden:1998})
show that even this simple analysis can give interesting 
results, the method we present can easily be generalized to include
variable sequences and other more complicated patterns.
\vc
The computational method we propose has two main steps: first the open 
reading frames (ORFs) of an eukaryote genome are grouped in
(overlapping) sets based on words  that are
overrepresented 
in their upstream region, compared to their frequencies in  the
reference sample 
made of all the 
upstream regions of the whole genome. 
Each set is labelled by a word.
Then for each of these sets the average
expression in one or more microarray experiments are compared to the
genome-wide average: if a statistically significant difference is
found, the word that labels the set is a candidate regulatory site
for the genes in the set, either enhancing or inhibiting their
expression.
\vc 
It is worth stressing that the grouping of the genes into sets depends 
only on the upstream sequences and not on the microarray experiment
considered: It needs to be done only once for each organism, 
and can then be used to analyse an arbitrary number of microarray
experiments. It is precisely  this fact that should allow the
extension of the method to patterns 
more complex than fixed sequences, while keeping the required
computational resources within
reasonable limits.
\vc
{\bf Constructing the sets}
\nopagebreak
\vc
We consider the upstream region of each open reading frame (ORF), and
we fix the maximum length $K$ of the upstream sequence to be
considered. The choice of $K$ depends on  the
typical location of most regulatory sites: in general $K$ is a number
between several hundred and a few thousand. For each ORF $g$, the
actual length of the sequence we consider is $K_g$ defined as the
minimum between $K$ and the available upstream sequence before the
coding region of the previous gene. 
\vc
For each word $w$ of length $l$ ($6\leq l \leq 8$ in this study),
and for each ORF $g$ we compute the number $m_g(w)$ of occurrences of
$w$ in the upstream region of $g$.
Non palindromic words are counted on both strands: therefore we define 
the effective number of occurrences  $n_g(w)$ as
\ba
n_g(w)&=&m_g(w)+m_g(\tilde{w}) \qquad {\rm if}\  w\ne\tilde{w}\\
n_g(w)&=&m_g(w)\qquad {\rm if}\  w=\tilde{w}
\ea
where $\tilde{w}$ is the reverse complement of $w$.
\vc
 We define the global frequency $p(w)$ of each word $w$ as
 \be
 p(w)=\frac{\sum_g n_g(w)}{\sum_g L_g(w)}
 \ee
where, in order to count correctly the available space for palindromic 
and non palindromic words,
 \ba
L_g(w)&=&{2(K_g-l+1)}\qquad {\rm if}\  w\ne\tilde{w}\\
L_g(w)&=&{(K_g-l+1)}\qquad {\rm if}\  w=\tilde{w}
\ea
$p(w)$ is therefore the frequency with which the word $w$ appears in
the upstream regions of the whole genome: it is the ``background
frequency'' against which occurrences in the upstream regions of the
individual genes are compared to determine which words are overrepresented.
\vc
For each ORF $g$ and each word $w$ we compute the probability
$b_g(w)$ of finding $n_g(w)$ or 
more occurrences of $w$ based on the global frequency $p(w)$:
\be
b_g(w)=\sum_{n=n_g(w)}^{L_g(w)} \left( L_g(w) \atop n\right)
p(w)^n \left[ 1-p(w)\right]^{L_g(w)-n}
\ee
\vc
We define a maximum probability $P$, depending in general on the
length $l$ of the words under consideration, 
and consider, for each $w$, the set 
\be 
S(w)=\{ g:\ b_g(w)<P\}
\ee 
of the ORFs in which the word $w$ is
overepresented compared to the frequency of $w$ in the upstream
regions of the whole genome. That is, $w$ is considered
overrepresented in the upstream region of $g$ if the probability of
finding  $n_g(w)$ or more instances of $w$ based on the global
frequency is less than $P$.
\vc
This completes the construction of the sets $S(w)$. Two free parameters 
have to be fixed: the length $K$ of the upstream region to be considered
and the probability cutoff $P$ for each length $l$ of words 
considered. A result in Ref. \cite{vanHelden:1998} suggests suitable
choices of these two numbers: the authors list the 34 ORFs of {\it
S. cerevisiae} that have 3 or more occurrences of the word GATAAG
in their 500 bp upstream region. 23 out of these 34 ORFs correspond to a
gene with known function, and 20 out of these 23 are regulated by
nitrogen. This result suggests to choose $K=500$ for the
upstream length, and a value of the probability cutoff such that three
or more instances of GATAAG in the 500 bp upstream region of an ORF are
considered significant. 
Any choice of $P$ between 0.018 and 0.1 would
satisfy this criterion, and we chose $P=0.02$.
Tentatively, we kept the same value of $P$
for all values of $l$. With this choice, the number of instances of a
word that are 
necessary to be considered overrepresented in a 500bp upstream
sequence  can be as high as six for
common 6-letter words and as low as one for rare 8-letter words. In
particular, our set $S(GATAAG)$ almost\footnote{Our set is smaller
that the one reported in Ref. \cite{vanHelden:1998} because we do not
allow the upstream sequence to overlap with the previous gene: this
eliminates 7 genes form the set.} coincides with the one
discussed in \cite{vanHelden:1998}. However the word GATAAG will not
turn out to be significant in our study.
\vc
As noted above, it would be natural to make the probability cutoff $P$ 
depend on the word length, simply because the number of possible words 
increases with their length: For example one could take the cutoff for
each word length 
to be inversely proportional to the number of independent words of
such length. However it turns out that this procedure tends to
construct sets that are less significant when tested for correlation
with expression. Therefore we chose to fix the cutoff at 0.02 for all
word lengths. It is important to keep in mind that no statistical
significance whatsoever is attributed to the sets {\it per se}: The
only sets that are retained at the end of the analysis are the ones
that show significant correlation with expression. Therefore the
choice of the cutoff in the construction of the sets can be based on
such a pragmatic approach without jeopardizing the statistical
relevance of the final result.
\vc
{\bf Studying the average expression level in each set}
\nopagebreak
\vc
The second step of our procedure consists in studying, for each set
$S(w)$ defined as above, the expression profiles of the ORFs belonging
to $S(w)$ in DNA microarray experiments. The idea is that if the
average expression profile in the set $S(w)$ for a certain experiment
is significantly different from the average expression for the same
experiment computed on the whole genome, then it is likely that some
of the ORFs in $S(w)$ are coregulated {\it and} that the word $w$
is a binding site for the common regulating factor. 
\vc
To look for such instances we consider the gene expression profiles
during the {\it diauxic shift}, {\it i.e.} the 
metabolic shift from fermentation to respiration, as
measured with DNA microarrays techniques in Ref. \cite{DeRisi:1997}.
In the experiment gene expression levels were measured for virtually
all the genes of {\it S. Cerevisiae} at seven time-points while such
metabolic shift took place. The experimental
results are publicly available from the web supplement to
Ref. \cite{DeRisi:1997}.   
\vc
We considered each time-point as a single experiment, and for each
gene $g$ we defined the quantity $r_g(i)\ (1=1,\dots,7)$ as the
$\log_2$ of the ratio between the mRNA levels for the gene $g$ at
time-point $i$ and the initial mRNA level. Therefore e.g. $r_g(i)=1$ means
a two-fold increase in expression at timepoint $i$ compared to initial 
expression.
\vc
For each time-point $i$ we computed the genome-wide average expression 
$R(i)$ and its standard deviation $\sigma(i)$. 
These are reported in Tab. 1, where $N(i)$ is the number of genes with 
available expression value for each timepoint.
Then for each word $w$ we
compute the average expression in the subset of $S_w$ given by the
genes for which an experimental result is available at timepoint $i$
(in most cases this coincides with $S_w$):
\be
R_w(i)=\frac{1}{N(i,w)} \sum_{g\in S_w} r_g(i)
\ee
where $N(i,w)$ is the number of ORFs in $S_w$ for which an
experimental result at timepoint $i$ is available, and the difference
\be
\Delta R_w(i)=R_w(i)-R(i)
\ee
$\Delta R_w(i)$ is the discrepancy between the genome-wide average
expression at time-point $i$ and the average expression at the same
time-point of the ORFs that share an abundance of the word $w$ in
their upstream region. A significance index ${\rm sig}(i,w)$ is
defined as
\be
{\rm sig}(i,w)=\frac{\Delta R_w(i)}{\sigma(i)}\sqrt{N(i,w)}
\label{sig}
\ee
and the word $w$ is considered significantly correlated with
expression at time point $i$ if
\be
|{\rm sig}(i,w)|>\Lambda
\ee
In this work we chose $\Lambda=6$: this means that we consider
meaningful a deviation of $R_w(i)$ by six s.d.'s from its expected
value. The sign of ${\rm sig}(i,w)$ indicates whether $w$ acts as an
enhancer or an inhibitor of gene expression.
\vu
{\bf \large Results and discussion}
\nopagebreak
\vc
We found a total of 29 words of length between 6 and 8 above our
significance threshold $|{\rm sig}|>6$. Most of them are related to known
regulatory motifs; two words turned out to be false positives due to
the presence, in their sets, of families of identical ORF's. Finally,
one word does not match any known motif and is a candidate new binding 
site.
\vc
The comparison between our significant words and 
known motifs was performed using the database of regulatory
motifs made publicly available by the authors of
Ref. \cite{Pilpel:2001}, and the CompareACE software \cite{Hughes:2000} 
available from the same web source. This package allowed us to compute 
the Pearson correlation coefficient of the best alignment between each 
of our significant words and each known regulatory motif (expressed as 
a set of nucleotide frequencies).
\vc
We used the following criterion to associate our significant words to
known motifs: a motif is considered as identified if at least one
significant word scores better than 0.8 when compared to it. 
A probability value for this choice of the cutoff can be estimated to
be a few percent: out of all the 2080 independent 6-letter words, 66
(that is 3.17\%) score better than 0.8 with at least one motif. For 7- 
and 8-letter words we have respectively 2.21\% and 1.51\%.
Once a
motif has been identified, all words which score best with the motif
are attributed to it, independently of the score, provided their
expression pattern is consistent with the word(s) scoring better than
0.8. 
\vc
{\bf PAC and RRPE motifs}
\nopagebreak
\vc
Nine significant words can be associated to the PAC
motif \cite{Dequard-Chablat:1991,Tavazoie:1999,Hughes:2000},  all of
them with rather high scores. They are shown in Tab. 2, where, as in
all the following tables, 
significativity indices are shown only for those timepoints where
they exceed our threshold $|{\rm sig}|>6$. 
Given the perfect alignment of these words, it is not surprising that
these sets largely overlap each other: The
union af all the nine sets contains a total of 96 genes. 
As an example, in Fig. 1 we show the average expression for the genes
associated with the word GATGAG as a function of the time, compared to 
the average expression computed over the whole genome. Fig. 2 shows
the significance index for the same set.
In Tab. 3 we show the set of 24 genes associated to the 
word GATGAG, together with their expression profiles.
\begin{figure}
\centering
{\includegraphics[width=14.cm]{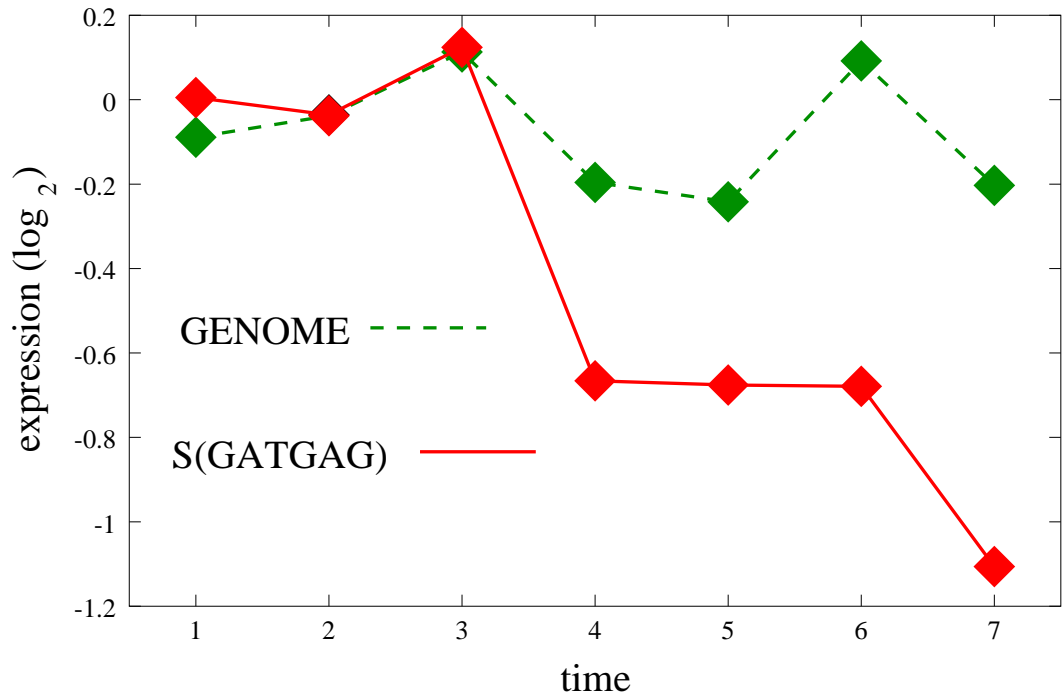}\caption{\it Expression of the
genes in the set S(GATGAG): The average expression of the genes in the
set 
(solid red line) are compared
to the genome-wide average expression (dashed green line) 
at the seven time points of the
diauxic shift experiment. The expression data are the $\log_2$ of the
ratio between mRNA levels at each timepoint and the initial mRNA
level.}} 
\end{figure}
\begin{figure}
\centering
{\includegraphics[width=14.cm]{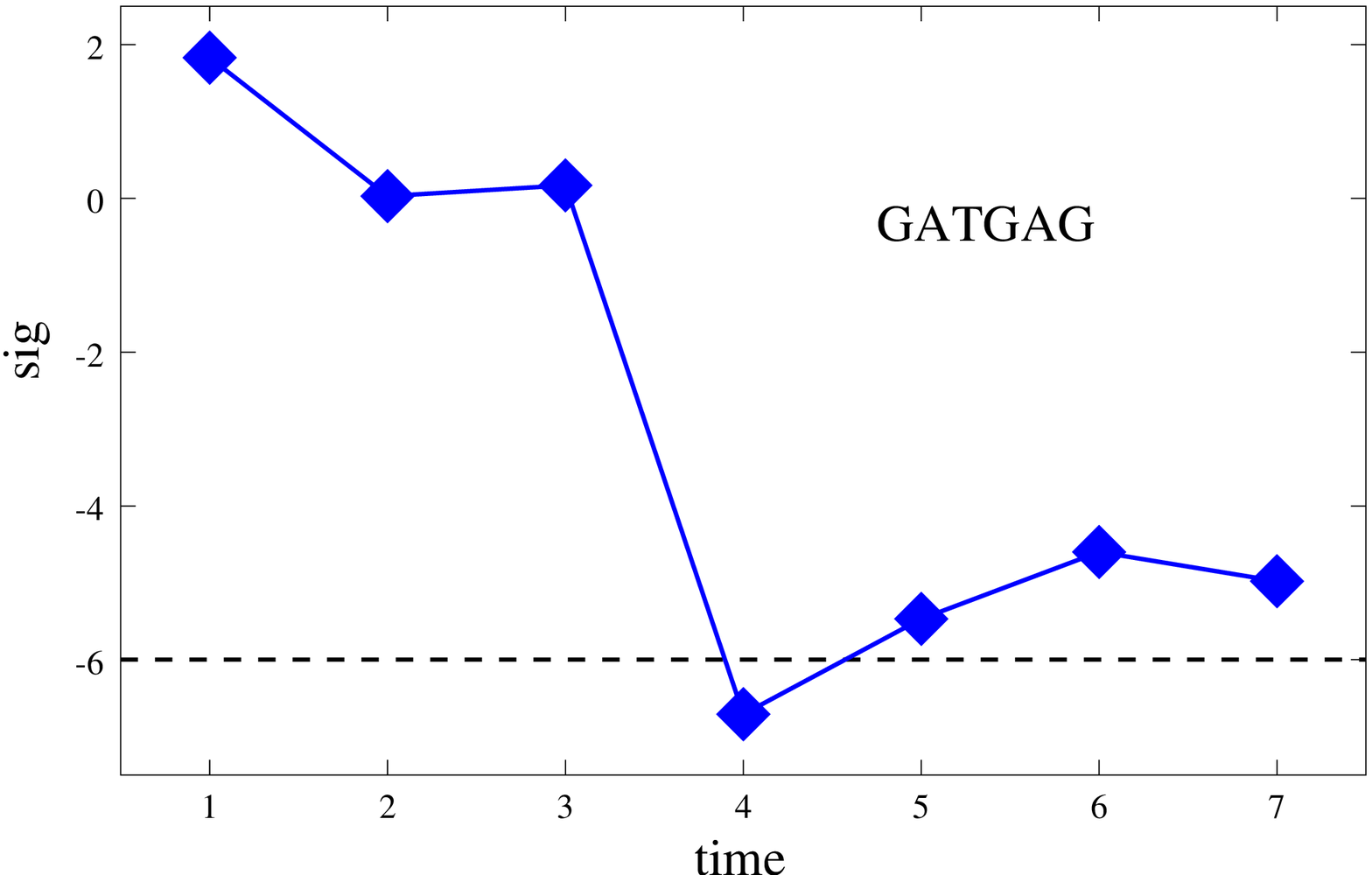}\caption{\it Statistical
significance ${\rm sig}(i,w)$ as defined in 
Eq.(\ref{sig}) for the word $w={\rm GATGAG}$ and timepoints
$i=1,\dots,7$ in the diauxic shift experiment. The dashed line is the
significance threshold $|{\rm sig}|=6$.}} 
\end{figure}
\vc
Two words can be associated with confidence
to the motif RRPE \cite{Tavazoie:1999, Hughes:2000}, and are shown in Tab. 4.
The union of the two sets contains 76 genes.
We see that genes containing the motifs PAC and RRPE are {\em repressed} at the
late stage of the diauxic shift compared to the early stages.
This result is in agreement with the expression
coherence score data available from the web supplement to
Ref. \cite{Pilpel:2001}: There one can see that (1) of all known
regulatory motifs, PAC and RRPE show the highest expression coherence
for the diauxic shift and (2) {\it viceversa}, of the eight
experimental conditions considered in Ref. \cite{Pilpel:2001}, the
diauxic shift is the one in which both the PAC and RRPE motif show the
highest expression coherence score.  
\vc
{\bf STRE and MIG1 motifs}
\nopagebreak
\vc
A total of ten significant words can be associated to the motifs
STRE \cite{Kobayashi:1993, Martinez-Pastor:1996}  and 
MIG1 \cite{Nehlin:1990,Ostling:1996}. 
It is well known that these play an important role in glucose
repression (see {\it e.g.} \cite{DeRisi:1997, Johnston:1999} and
references therein).
Most of these words show comparable scores for the two
motifs (due to their similarity) so we decided to show them
together in Tab. 5 which 
shows the two scores for each word. A total of 212 genes belong to the 
union of all these sets.
\vc
{\bf The UME6 motif}
\nopagebreak
\vc
Two words are associated to the known UME6 motif, a.k.a. URS1
\cite{Sumrada:1987, Gailus-Durner:1997}, known to be a pleiotropic
regulator implicated in glucose repression \cite{Kratzer:1997}.  They
are shown in Tab. 6. The two sets do not overlap, so that a total of 
56 genes are associated to this motif.
\vc
{\bf Other significant words}
\nopagebreak
\vc
Three words, shown in Tab. 7, are of uncertain status: for the first
one, the set S(ACTTTC) contains only 2 genes, making the statistical
significance of the result questionable. The word CCCCTGAA scores best
with the PDR motif (0.58): given the low significance of this score,
and the fact that PDR does not seem to be relevant for any other word, 
this is most likely accidental. The word should probably be considered 
as belonging to the STRE/MIG1 motif (the scores are STRE: 0.46, MIG1: 0.49).
Finally the word GCCCCTGA scores best with UME6 (0.55), but its expression pattern is 
more similar to the STRE/MIG1 motifs (scores: STRE:0.44, MIG1: 0.46).
\newpage\noindent
{\bf False positives due to families of identical or nearly identical ORF's}
\vc
The genome of {\it S. cerevisiae} contains a few families of genes
whose coding {\it and} upstream regions are identical or nearly
identical. Consider for example the COS1 gene (YNL336W): the seven
genes COS2-COS8 have both coding sequence and 500kb upstream sequence
coinciding better than 80\% with the COS1 sequence. Therefore if the
upstream sequence of COS1 contains overerpresented words, they will likely
appear in all of the upstream regions. On the other hand, the
expression profiles of all the genes in the family will be the same
when measured by a microarray experiment, simply because the
experimental apparatus cannot distinguish between the mRNA produced by 
the various members of the family, due to cross-hybridization between
their mRNA. Therefore all of the genes of the
family are likely to occur in the sets of the words that are
overrepresented in their upstream region, and even a small
deviation from the genome-averaged expression acquires a statistical
significance. 
\vc
We found two instances of this in our analysis: the words GACGTAGC and 
GGTCGCAC appear to be associated to significant enhancement of the
corresponding sets of genes at late timepoints in the diauxic shift:
however the two sets contain respectively seven out of eight and all of the 
COS1-COS8 genes. Since the COS genes are mildly overexpressed, this creates a
false statistical significance. When one corrects for this, by keeping 
only one representative of the family, the statistical significance of 
the two sets disappears.
\vc
{\bf A candidate new motif}
\nopagebreak
\vc
Finally, the word ATAAGGG/CCCTTAT is a candidate new
binding site, since it does not have good comparison scores with any 
of the known motifs. 
It scores best with the AFT1 motif, with a 0.52 score which is
practically meaningless since 
84.9\% of all independent 7-letter words score the same or better with at 
least one motif.
It is associated with 13 genes, as shown in
Tab. 8, which are overexpressed at late timepoints. The average
expression levels for the set and the significance index are shown as
a function of time in Figs. 3 and 4.
\begin{figure}
\centering
{\includegraphics[width=14.cm]{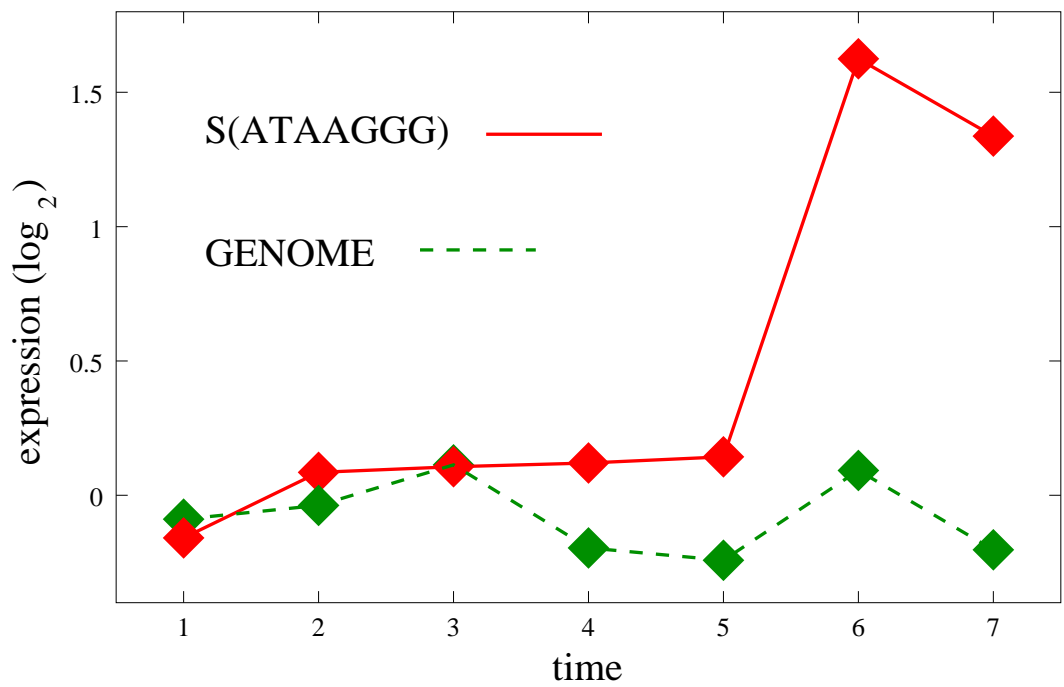}\caption{\it Expression of the
genes in the set S(ATAAGGG): Same as Fig. 1 for  our new candidate
regulatory motif.}} 
\end{figure}
\begin{figure}
\centering
{\includegraphics[width=14.cm]{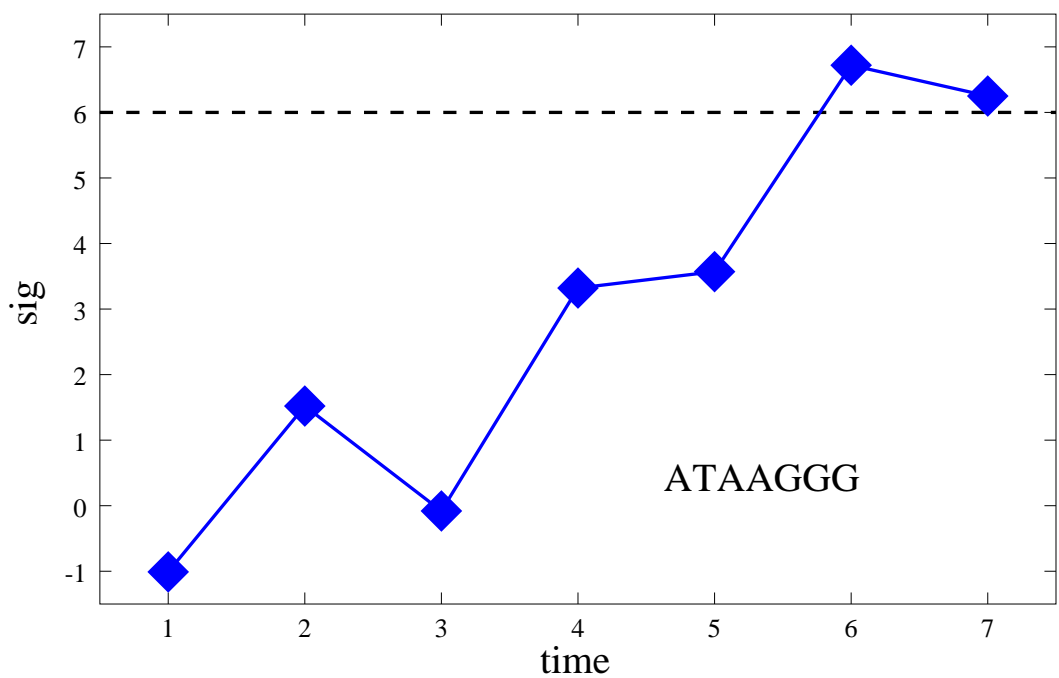}\caption{\it Statistical
significance of the set S(ATAAGGG): Same as Fig. 2 for the word
ATAAGGG.}}  
\end{figure}
\vc
{\bf Comparison with the results of Ref. \cite{Bussemaker:2001}}
\vc
As stated in the introduction, the method proposed in
Ref.\cite{Bussemaker:2001} also allows one to identify regulatory
motifs without any previous clustering of gene expression data: a
linear dependence  
of the logarithm of the expression levels on the number of repetitions 
of each regulatory motifs is postulated, and motifs are ranked
according to the reduction in $\chi^2$ obtained when such dependence
is subtracted from the experimental expression levels. Iteration of
the procedure produces a {\it model}, that is a set of relevant
regulatory motifs, for each expression data set.
\vc
In Ref. \cite{Bussemaker:2001} such a model is presented for the 14
min. time point in the $\alpha$-synchronized cell-cycle experiment of
Spellmann {\it et al.}, Ref. \cite{Spellmann:1998}. We used our
algorithm on the same data set to compare the findings.
Let us concentrate on the 7-letter words (the longest considered in 
\cite{Bussemaker:2001}). We found 9 significant words,
reported in Tab. 9. Of these, five coincide with or are very similar
to words found by the authors of  Ref.\cite{Bussemaker:2001} (see
their Tab. 2). The remaining four (AGGCTAA, GGCTAAG, GCTAAGC and
CTAAGCG, whose similarity clearly suggests the existence of a longer
motif) are of particular interest for the purpose of comparing the two 
methods: If one looks at the dependence of the expression levels on
the number of occurrences of these words in the 500 bp upstream
region, one clearly sees the existence of an activation threshold (see 
Fig. 5, where such dependence is shown for GGCTAAG). On the other hand, by
looking at these data one hardly expects a significant reduction in
$\chi^2$ when trying to describe this dependence with a straight line.
This should be compared to the same dependence for the word AAAATTT,
shown in Fig. 6, which is found by {\it both} algorithms. 
On the other hand, there are two 7-word motifs found in
\cite{Bussemaker:2001} that do not pass our significativity threshold,
that is CCTCGAC and TAAACAA.
\begin{figure}
\centering
{\includegraphics[width=14.cm]{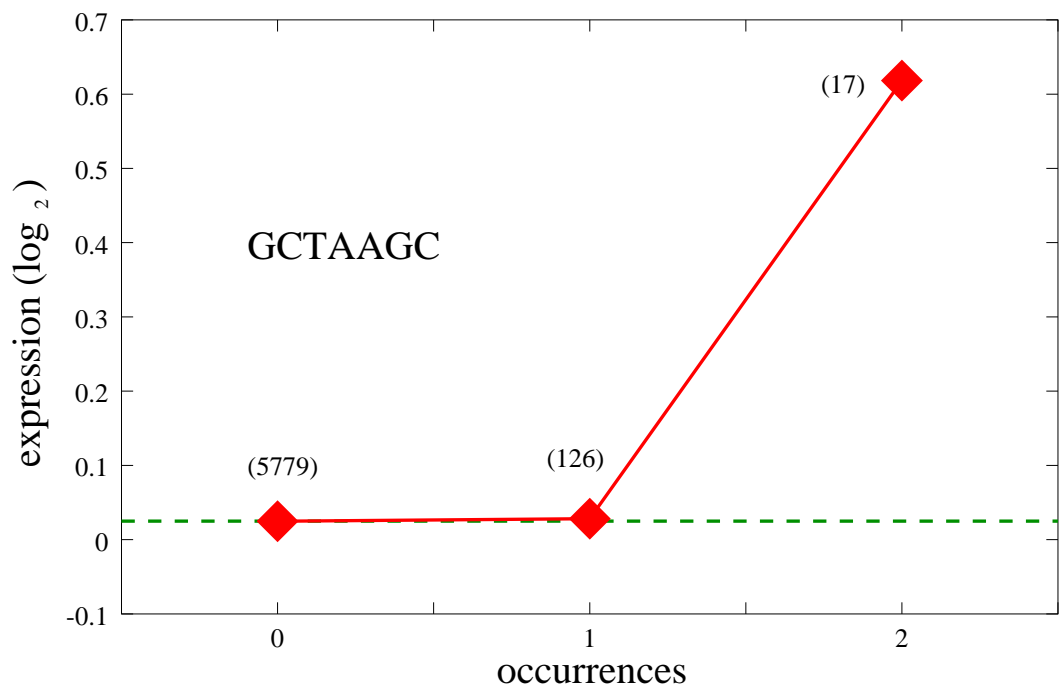}\caption{\it Expression as a
function of occurrences of the word GGCTAAG: The average expression of
genes presenting $n$ occurrences of the word  
GGCTAAG as a function of $n$ in the 14 min. time point of the 
$\alpha$-synchronized cell-cycle experiment of
Spellmann {\it et al.}, Ref. \cite{Spellmann:1998}. 
In parentheses is the number of genes with $n$ occurrences of 
GGCTAAG in the upstream region. The horizontal
line represents the average expression for the whole genome.}}
\end{figure}
\begin{figure}
\centering
{\includegraphics[width=14.cm]{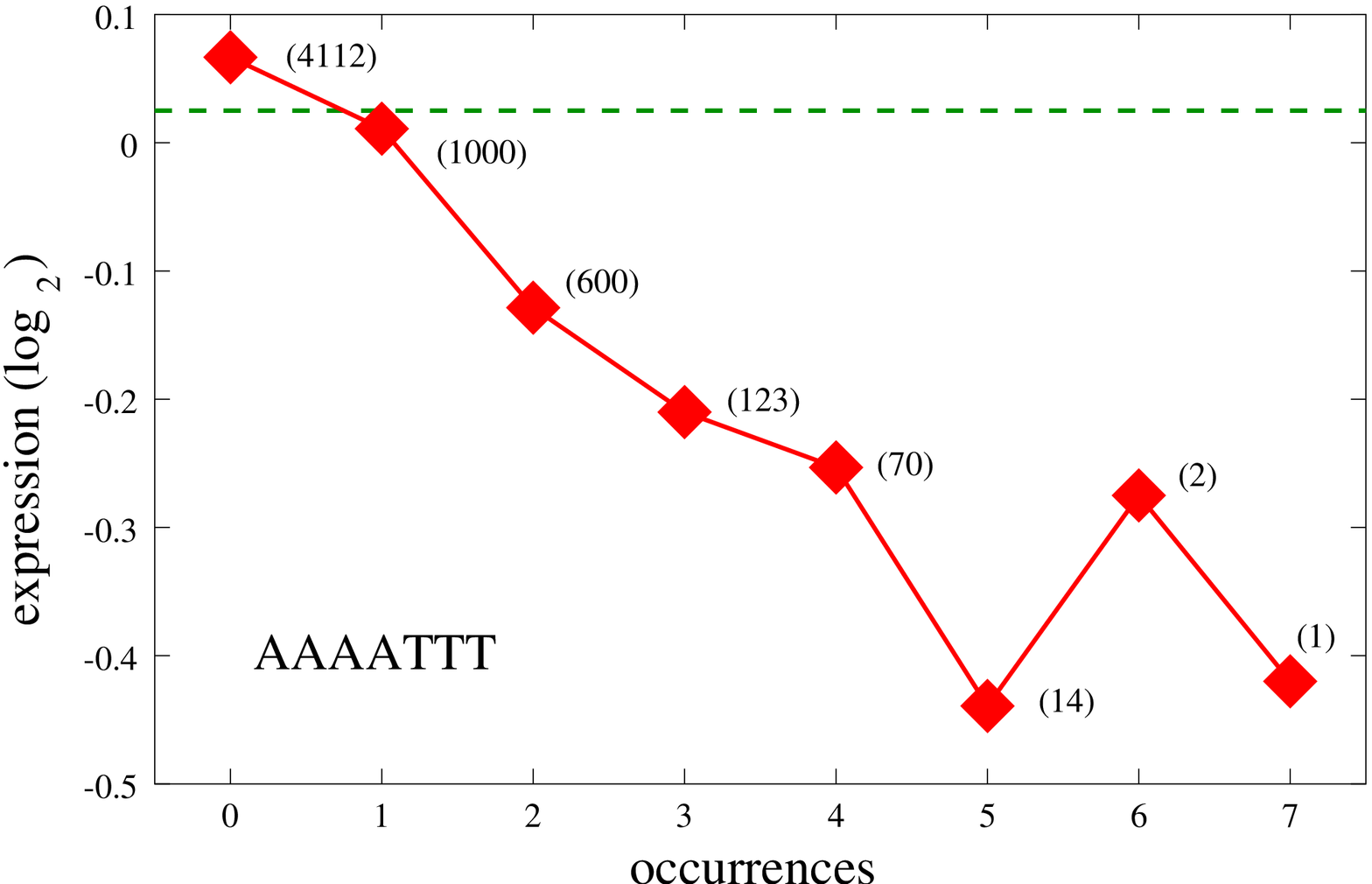}\caption{\it Expression as a
function of occurrences of the word
AAAATTT: Same as Fig. 5 for AAAATTT.}}
\end{figure}
\vc
We can conclude that the two methods tend to find motifs with a
different effect on gene expression: probably the best results can be
obtained by using them both on the same data set.
\vu
{\bf \large Conclusions}
\nopagebreak
\vc
We have presented a new computational method to identify regulatory
motifs in eukaryotes, suitable to identify those  motifs that are
effective when repeated many times in the upstream sequence of a
gene. The main feature that differentiates our method from existing
algorithms for motif  discovery is the fact that genes are
grouped {\it a priori} based on similarities in their upstream
sequences. 
\vc
Most of the significant words the algorithm finds can be associated to 
five known regulatory motifs: This fact consitutes a strong validation
of the method. Three of them (STRE, MIG1 and UME6) were previously  known to be
implicated in glucose suppression, 
while the fact that PAC and RRPE sites are relevant to regulation
during  the diauxic shift is in agreement with expression coherence
data as reported in the web supplement to Ref. \cite{Pilpel:2001}.
One of the significant words we find (ATAAGGG) cannot be
identified with any known motif, and is a candidate new binding site.   
\vc
It is easy, at least in principle, to extend the method to a larger
class of regulatory sites. According to our knowledge of gene
regulation, this should be done at least in two directions: (1) the
analysis should not be restricted to fixed sequences, but extended to motifs
with controlled variability; in particular the extension to spaced
dyads \cite{vanHelden:2000} should be straightforward; (2) the {\it
combinatorial} analysis of binding sites \cite{Pilpel:2001} could also 
be performed along the same lines, that is first grouping genes
according to which combinations of motifs appear in their upstream
region, and then analysing expression profiles within each group.   
\vu
{\bf Acknowledgement} 
\nopagebreak
\vc
Upstream sequences were downloaded, and many
cross-checks were performed, using the impressive collection of packages
available from the Regulatory Sequence Analysis Tools (RSAT)
\cite{vanHelden:2000b} 
at \\*
http://www.ucmb.ulb.ac.be/bioinformatics/rsa-tools/  or\\*
http://embnet.cifn.unam.mx/rsa-tools/.
\vc

\begin{table}[t]
\begin{center}
\begin{tabular}{|c|r|r|r|}
\hline
$i$&$N(i)$&$R(i)$&$\sigma(i)$\\
\hline
       1      &6082  &-0.0888  & 0.2509\\
\hline
       2      &6054  &-0.0378  & 0.2801\\
\hline
       3      &6020  & 0.1132  & 0.3152\\
\hline
       4      &6071  &-0.1957  & 0.3433\\
\hline
       5      &6058  &-0.2423  & 0.3890\\
\hline
       6      &6084  & 0.09244  & 0.8226\\
\hline
       7      &6021  &-0.2028  & 0.8886\\

\hline
\end{tabular}
\end{center}
\caption {\sl Number of data, average and standard deviation
  for the 7 timepoints.}
\end{table}
\begin{table}[h]
\begin{center}
\begin{tabular}{|c@{}c@{}c@{}c@{}c@{}c@{}c@{}c@{}c@{}c@{}c@{}c@{}|c|ccccccc|c|}
\hline
\multicolumn{12}{|c|}{word} &genes&\multicolumn{7}{c|}{timepoints}&score\\
\multicolumn{12}{|c|}{}& &1&2&3&4&5&6&7&\\
\hline
\hline
 & & & &G&A&T&G&A&G& & &24&---&---&---&-6.70&---&---&---&1.00\\
 & & & &G&A&T&G&A&G&A&T&35&---&---&---&-8.20&-6.26&-6.18&-7.86&0.94\\
 & & & &G&A&T&G&A&G&A& &26&---&---&---&-7.06&---&---&-6.64&0.93\\
 & &G&A&G&A&T&G&A&G& & &36&---&---&---&-6.96&---&---&-6.50&0.92\\
 & & &A&G&A&T&G&A&G& & &33&---&---&---&-6.17&---&---&-6.44&0.91\\
 & &G&A&G&A&T&G&A& & & &42&---&---&---&-6.20&---&---&---&0.83\\
A&T&G&A&G&A&T&G& & & & &32&---&---&---&-6.96&---&---&-6.33&0.80\\
 & &G&A&G&A&T&G& & & & &31&---&---&---&-6.42&---&---&---&0.75\\
 &T&G&A&G&A&T&G& & & & &47&---&---&---&-6.26&---&---&-6.10&0.70\\
\hline
\end{tabular}
\end{center}
\caption {\sl Significant words related to the PAC motif.}
\end{table}
\begin{table}[h]
\begin{center}
\begin{tabular}{|l|l|rrrrrrr|}
\hline
ORF&gene&\multicolumn{7}{c|}{timepoints}\\
   &    &1&2&3&4&5&6&7\\
\hline
YBL054W&              &0.21 &-0.01 &-0.18 &-1.56 &-1.25 &-0.79&-1.47\\          
YCL059C&  KRR1             &0.36  &0.06  &0.45 &-0.69 &-0.71 &-0.34 &-1.69\\          
YDL063C&             &-0.03 &-0.03 &-0.27 &-0.92 &-1.06 &-1.51 &-2.06\\          
YDL153C&  SAS10            &0.41  &0.19  &0.36 &-0.76 &-0.97 &-1.43 &-1.79\\          
YDR365C&              &0.03  &0.06  &0.21 &-0.38 &-0.62 &-1.64 &-1.94\\          
YGR022C&             &-0.17 &-0.06  &0.14  &0.04 &-0.15  &0.54  &0.86\\          
YGR102C&             &-0.23 &-0.23 &-0.07 &-0.32  &0.03  &1.43  &0.84\\          
YGR103W&  NOP7            &0.15 &-0.06  &0.32 &-0.92 &-1.09 &-1.64 &-2.56\\          
YGR128C&              &0.30  &0.26  &0.38 &-0.81 &-0.76 &-0.89 &-1.47\\          
YGR129W&  SYF2           &-0.18 &-0.54  &0.11 &-0.12 &-0.23  &0.74  &0.14\\          
YGR145W&              &0.00 &-0.23  &0.25 &-0.92 &-1.09 &-1.69 &-2.18\\          
YJL033W&  HCA4           &-0.06  &0.01  &0.21 &-0.94 &-0.36 &-0.67 &-0.62\\          
YKL078W&             &-0.04 &-0.01  &0.04 &-1.12 &-0.97 &-0.71 &-1.89\\          
YKL172W& EBP2             &0.12  &0.21  &0.30 &-0.74 &-0.56 &-0.42 &-1.40\\          
YLR276C&  DBP9            &0.03  &0.14  &0.32 &-0.62 &-0.86 &-0.67 &-1.64\\          
YLR401C&             &-0.06 &-0.07  &0.07 &-0.71 &-0.71 &-0.84 &-1.03\\          
YLR402W&             &-0.18 &-0.23 &-0.30 &-0.47 &-0.51 &-0.20 &-0.27\\          
YML123C&   PHO84      &0.50  &0.50  &0.54 &-0.56 &-0.67 &-2.32 &-1.69\\          
YNL061W&   NOP2      &-0.03 &-0.51 &-0.42 &-1.29 &-1.36 &-2.25  &0.01\\          
YNL062C&   GCD10     &-0.10  &0.00  &0.01 &-0.47 &-0.64 &-1.12 &-1.06\\          
YOL141W&   PPM2          &-0.10  &0.01  &0.24 &-0.84 &-0.54  &0.04 &-0.20\\          
YPL068C&             &-0.60 &-0.10 &-0.18 &-0.84 &-1.09  &0.08 &-0.89\\          
YPR112C&   MRD1          &-0.17 &-0.23 &-0.17 &-0.54 &-0.62 &-1.12 &-1.51\\          
YPR113W&   PIS1      &-0.04  &0.00  &0.62  &0.52  &0.56  &1.12 &-1.03\\          
\hline
\multicolumn{2}{|c|}{set average}
&0.005&-0.036&0.124&-0.666&-0.676&-0.679&-1.106\\
\multicolumn{2}{|c|}{genome average}
&-0.089&-0.038&0.113&-0.196&-0.242&0.092&-0.203\\
\multicolumn{2}{|c|}{significance}
&1.83&0.03&0.17&-6.71&-5.47&-4.60&-4.98\\
\hline
\end{tabular}
\end{center}
\caption 
{\sl The ORFs in the set S(GATGAG) with their expression profiles.}
\end{table}
\begin{table}[th]
\begin{center}
\begin{tabular}{|c@{}c@{}c@{}c@{}c@{}c@{}c@{}c@{}|c|ccccccc|c|}
\hline
\multicolumn{8}{|c|}{word} &genes&\multicolumn{7}{c|}{timepoints}&score\\
\multicolumn{8}{|c|}{}& &1&2&3&4&5&6&7&\\
\hline
\hline
A&A&A&A&T&T&T& &50&---&---&---&---&---&-7.90&-8.58&0.91\\
A&A&A&A&T&T&T&T&62&---&---&---&-6.59&---&-8.73&-10.26&0.89\\
\hline
\end{tabular}
\end{center}
\caption {\sl Significant words related to the RRPE  motif.}
\end{table}
\begin{table}[h]
\begin{center}
\begin{tabular}{|c@{}c@{}c@{}c@{}c@{}c@{}c@{}c@{}c@{}c@{}c|c|ccccccc|cc|}
\hline
\multicolumn{11}{|c|}{word}
&genes&\multicolumn{7}{c|}{timepoints}&\multicolumn{2}{c|}{score}\\ 
\multicolumn{11}{|c|}{}& &1&2&3&4&5&6&7&STRE&MIG1\\
\hline
\hline
C&C&A&C&C&C&C&C& &      &     &35&---&---&---&---&---&6.39&---&0.82&0.53\\
 &C&C&C&C&C&C&C&T&      &     &28&---&---&---&---&---&6.01&---&0.79&0.71\\
 & & & &C&C&C&C&T&G     &     &28&---&---&---&7.06&6.09&7.00&---&0.59&0.54\\
 &C&A&G&C&C&C&C&T&      &     &23&---&---&---&---&---&6.42&---&0.59&0.42\\
 & & &G&C&C&C&C&T&$^*$  &     &40&---&---&---&---&---&7.05&---&0.59&0.56\\
 & &G&C&C&C&C&C&T&G     &$^*$ &17&---&---&---&---&6.07&---&---&0.47&0.46\\
\hline
 & &T&A&C&C&C&C& &      &     &25&---&---&---&---&---&6.09&---&0.55&0.85\\
 & &C&C&C&C&C&C& &      &     &56&---&---&---&---&6.48&6.55&6.10&0.72&0.80\\
 & & &A&C&C&C&C&T&      &     &29&---&---&---&---&---&7.42&---&0.63&0.65\\
 & &G&G&C&C&C&C& &      &     &16&---&---&---&---&6.71&---&---&0.52&0.56\\
\hline
\end{tabular}
\end{center}
\caption {\sl Significant words related to the STRE and MIG1
motifs.The words marked * actually score better with the variant STRE' motif
(0.60 and 0.55 respectively).}
\end{table}
\begin{table}[th]
\begin{center}
\begin{tabular}{|c@{}c@{}c@{}c@{}c@{}c@{}c@{}c@{}c@{}c|c|ccccccc|c|}
\hline
\multicolumn{10}{|c|}{word} &genes&\multicolumn{7}{c|}{timepoints}&score\\
\multicolumn{10}{|c|}{}& &1&2&3&4&5&6&7&\\
\hline
 &G&C&C&G&C&C& & &\qquad &27&---&---&---&---&---&---&6.03&0.82\\
A&G&C&C&G&C&G&C& & &29&---&---&---&---&---&---&6.63&0.60\\
\hline
\end{tabular}
\end{center}
\caption {\sl Significant words related to the UME6 motif.}
\end{table}
\begin{table}[th]
\begin{center}
\begin{tabular}{|c@{}c@{}c@{}c@{}c@{}c@{}c@{}c@{}c@{}c@{}|c|ccccccc|}
\hline
\multicolumn{10}{|c|}{word} &genes&\multicolumn{7}{c|}{timepoints}\\
\multicolumn{10}{|c|}{}& &1&2&3&4&5&6&7\\
\hline
A&C&T&T&T&C& & & & &2 &---&---&---&6.20&---&---&---\\
\hline
 &C&C&C&C&T&G&A&A& &42&---&---&---&6.50&---&---&---\\
G&C&C&C&C&T&G&A& & &22&---&---&---&6.90&---&---&---\\
\hline
\end{tabular}
\end{center}
\caption {\sl Significant words of uncertain attribution.}
\end{table}
\begin{table}[th]
\begin{center}
\begin{tabular}{|l|l|rrrrrrr|}
\hline
ORF&gene&\multicolumn{7}{c|}{timepoints}\\
   &    &1&2&3&4&5&6&7\\
\hline
YBR072W&   HSP26     &-0.01  &0.40  &0.36  &1.00  &1.43  &3.47  &2.84\\         
YDL133W&             &-0.04  &0.32 &-0.34 &-0.25 &-0.56 &-0.22 &-0.32\\          
YDL204W&             &-0.36  &0.92 &-0.51  &0.26  &0.08  &4.05  &3.06\\          
YIL136W&   OM45      &-0.97 &-0.27  &0.21 &-0.25  &1.32  &3.47  &1.79\\          
YLR163C&   MAS1       &0.04 &-0.01  &0.11 &-0.01  &0.08  &0.30 &-0.03\\          
YLR164W&             &-0.30  &N/A  &-0.27  &0.06 &-0.18  &2.19  &1.69\\          
YLR453C&   RIF2          &-0.07 &-0.27  &0.32 &-0.01 &-0.71  &0.69  &0.08\\          
YML127W&   RSC9           &0.01  &0.14  &0.08 &-0.18 &-0.27 &-0.30 &-1.06\\          
YML128C&   MSC1          &-0.12  &0.20  &0.97  &1.56  &1.36  &4.32  &3.47\\          
YNL117W&   MLS1      &-0.30 &-0.04  &0.71 &-0.30 &-0.27  &0.76  &3.18\\          
YPR025C&   CCL1      &-0.18 &-0.36 &-0.30 &-0.25 &-0.42  &0.36  &0.20\\          
YPR026W&   ATH1      &-0.06 &-0.04  &0.11  &0.20  &0.20  &0.60  &1.56 \\         
YPR172W&              &0.29  &0.03 &-0.07 &-0.27 &-0.20  &1.43  &0.92\\          
\hline
\multicolumn{2}{|c|}{set average}
&-0.159&0.085&0.106&0.120&0.143&1.625&1.337\\
\multicolumn{2}{|c|}{genome average}
&-0.089&-0.038&0.113&-0.196&-0.242&0.092&-0.203\\
\multicolumn{2}{|c|}{significance}
&-1.01&1.52&-0.08&3.32&3.57&6.72&6.25\\
\hline
\end{tabular}
\end{center}
\caption 
{\sl The ORFs in the set S(ATAAGGG) with their expression profiles.}
\end{table}
\begin{table}[th]
\begin{center}
\begin{tabular}{|c@{}c@{}c@{}c@{}c@{}c@{}c@{}c@{}c@{}c@{}c@{}c@{}|c|c|}
\hline
\multicolumn{12}{|c|}{word} &genes&{\rm sig}\\
\hline
\hline
 & & &A&A&A&A&T&T&T& & &50&-7.63\\
\hline
 & & &A&C&G&C&G&T&C& & &28&6.46\\
\hline
 & & &A&G&A&T&G&A&G& & &33&-6.96\\
 & & & &G&A&T&G&A&G&A& &25&-6.47\\
 & &G&A&G&A&T&G&A& & & &41&-6.60\\
\hline
 & &G&G&C&T&A&A&G& & & &17&7.30\\
 &A&G&G&C&T&A&A& & & & &22&6.65\\
 & & & &C&T&A&A&G&C&G& &16&6.89\\
 & & &G&C&T&A&A&G&C& & &17&6.77\\
\hline
\end{tabular}
\end{center}
\caption {\sl Significant 7-letter words for the 14-minute timepoint
in the $\alpha$-synchronized cell-cycle experiment}
\end{table}
\end{document}